\documentclass[12pt]{article}
\usepackage{color,amsmath,amssymb,amsthm,bm,epsf,epsfig,enumerate,alltt,latexsym,graphicx}
\usepackage{mathabx,textcomp}
\newcommand{\E}{\ensuremath{\mathrm{E}}}
\newcommand{\var}{\ensuremath{\mathrm{Var}}}
\newcommand{\pr}{\ensuremath{^{\prime}}}
\newcommand{\id}{\ensuremath{\mathrm{I}}}
\renewcommand{\P}{\ensuremath{\mathbf{P}\!}}
\newcommand{\N}{\ensuremath{\mathbf{N}}}
\renewcommand{\sp}{\ensuremath{\mathrm{sp}}}
\newcommand{\tr}{\ensuremath{\mathrm{tr}}}
\newcommand{\Diag}{\ensuremath{\mathrm{Diag}}}
\newcommand{\imp}{\ensuremath{\Longrightarrow}}
\newcommand{\pmi}{\ensuremath{\Longleftarrow}}
\newcommand{\mbk}{\ensuremath{\mathbb{K}}}
\newtheorem{propn}{Proposition}

\begin{document}
\begin{center}
\begin{Large}
{\bfseries Yates's and Other Sums of Squares}\end{Large}\\
\vspace{.5cm}by Lynn R. LaMotte\footnote{School of Public Health, LSU Health, New Orleans, LA,  {\tt llamot@lsuhsc.edu}}
\end{center}

\begin{abstract}
It is shown that the sum of squares by Yates's method of weighted squares of means is equivalent to numerator sums of squares formulated by other methods.  These relations are established first for hypotheses about fixed effects in a general linear model, in the process showing how Yates's method can be extended.  They are then illustrated in the unequal-subclass-numbers model for main effects and interaction effects of two factors.
\end{abstract}

\vspace{.5cm}\noindent{\sc Key Words}: ANOVA, Linear Models, Main Effects

\section{Introduction}
In a seminal paper, Yates (1934) described the ``method of weighted squares of means'' (MWSM) to obtain a numerator sum of squares for testing main effects of  factor A in an unbalanced  model that permits main effects of  factors A and B and their interaction effects.  He reasoned that, if $\bm{U}\sim\N(\mu\bm{1}_p, \sigma^2D)$, with $D=\Diag(1/w_i)$, all $w_i>0$, then, quoting Yates's equation (A),
\begin{eqnarray}
Q&=& (p-1)s^2 = w_1(u_1-\bar{u})^2 + w_2(u_2-\bar{u})^2 + \cdots \nonumber\\
&=& w_1u_1^2 + w_2u_2^2 + \cdots - (w_1+w_2+ \cdots)\bar{u}^2 \nonumber\\
& & \text{ where } \bar{u}=\frac{w_1u_1 + w_2u_2 + \cdots}{w_1+w_2+ \cdots}\label{Yates Q}
\end{eqnarray}
  ``provides an efficient estimate'' of $(p-1)\sigma^2$ from the realized value $\bm{u}=(u_1,\ldots,u_p)\pr$. In matrix terms, $Q$ can be expressed as
\begin{equation}Q=\bm{u}\pr(D^{-1}-D^{-1}\bm{1}(\bm{1}\pr D^{-1}\bm{1})^{-1}\bm{1}\pr D^{-1})\bm{u},\label{Q form}\end{equation}
where $\bm{u}=(u_1,\ldots, u_p)\pr$, $\bm{1}$ denotes a $p$-vector of ones, and $D = \Diag(1/w_i)$. 
	
The MWSM numerator sum of squares for A main effects comes from this expression upon substituting the ``marginal means of the subclass means'' for $u_i$, with corresponding substitutions for the diagonal entries of $D$.  

Yates proffered no further rationale.  He did not invoke a general approach or set of criteria.  For that reason, it is not clear how to develop $Q$ from basics, or how it is related to alternatively-developed sums of squares,  or how to extend it to other settings. 

Herr (1986) notes that in an earlier paper Yates (1933) ``indicated that [MWSM]  is a least squares procedure when he said that the variance for treatment in [MWSM] is `identical with the residual variance when constants representing [B main effects] and [AB interaction effects] are fitted' (p. 118).'' In Yates's usage, ``residual variance'' meant the increase in error sum of squares (SSE) upon deleting a set of terms from a full model.  That is termed here the restricted-model, full model difference in SSE, abbreviated RMFM.

The assertion that the MWSM sum of squares ``is a least squares procedure'' (an RMFM sum of squares) was not derived or proven in either of the Yates (1933, 1934) papers (Yates (1933, p. 118) says ``It can be shown ...''), nor did Herr (1986) give any mathematical justification for the assertion.  It is widely held, apparently, that this is true.  Perhaps this belief was based on direct experience, but direct proofs are hard to find.  Anderson and Bancroft (1952, p. 279) say that ``[the MWSM] provides exact tests of the main effects when interaction is present.'' They cite, among others, Snedecor and Cox (1935), who suggest (p. 246) that the MWSM ``is especially appropriate if the postulated population has equal subclass numbers. ... [I]f the method is applied to a sample with equal subclass numbers it yields exactly the same results as the standard method for such numbers; but if it is applied to a sample with proportional (but not equal) subclass numbers the results do not coincide with those obtained from the standard method for proportional numbers.''  It appears that these sources rely on examples and experience  rather than mathematical constructions. 

Searle (1971, p. 371) showed that the MWSM sum of squares tests equality of the A marginal means by showing that its  noncentrality parameter is 0 if and only if the marginal means are all equal.  That is apparent from (\ref{Q form}).
Searle, Speed, and Henderson (1981, Appendix B) related it directly to least squares by showing that it could be derived from the form $(G\pr\hat{\bm{\eta}})\pr[\var(G\pr\hat{\bm{\eta}})/\sigma^2]^-(G\pr\hat{\bm{\eta}})$ to test H$_0: G\pr\bm{\eta}=\bm{0}$ in the framework of the model $\bm{Y} \sim \N(\mbk\bm{\eta}, \sigma^2\id)$, where $\bm{\eta}$ is the vector of cell means, columns of  $G$  comprise a complete set of contrasts for the factor main effects in question, and $\hat{\bm{\eta}}$ is the vector of cell sample means. Searle (1987, p. 90) quoted the MWSM sum of squares directly as shown in Yates (1934) and then justified that the resulting $F$-statistic ``is a test statistic for'' the hypothesis of equal A marginal means because, if the marginal means are equal, then the MWSM sum of squares is distributed as proportional to a central chi-squared random variable. 

In 1934, a very positive feature of the MWSM sum of squares was that it was an explicit formula.  Today, with statistical computing packages, it should be possible to obtain an appropriate numerator sum of squares in any linear model for  hypotheses based on any set of estimable functions of the parameters of the mean vector. In models that involve effects of combinations of levels of multiple categorical factors, it is widely thought that the SAS Type III sum of squares (see SAS Institute 1978) is the correct numerator sum of squares for testable hypotheses. However, proofs are hard to find, and it is not always clear what the ``correct'' numerator sum of squares is.  

This topic, whether and how to test for main effects in models that do not exclude interaction effects, continues to generate much discussion.   See Searle (1994), Macnaughton (1998), Hector et al. (2010), Langsrud (2003), and Smith and Cribbie (2014). The books by Hocking (2013) and Khuri (2010) give detailed and comprehensive treatments of the topic.  Still, there is disagreement and some confusion on several points.  Those will not be resolved here.
 
The purpose of this paper is to describe several different approaches to constructing numerator sums of squares in a general linear model and to show that they all produce the same sum of squares for the same hypothesis.  One of the methods parallels Yates's rationale, and all of them produce the same MWSM sum of squares in the unbalanced two-factor analysis of variance model.

See Appendix \ref{defns} for definitions and notation used here.

\section{Numerator Sums of Squares for Estimable Functions}

In this section, four methods are presented that lead to numerator sums of squares.   It is shown that, for hypotheses about estimable functions, all give the same sum of squares.  Call them the geometric, restricted-model full model (RMFM), Pearson's chi-squared, and variance estimator heuristics.

Let $\bm{Y}$ denote an $n$-variate random variable, with realized value $\bm{y}$, that follows the model $\bm{Y}\sim \N(X\bm{\beta}, \sigma^2\id)$.  $X$ is a given $n\times k$ matrix of constants; $\bm{\beta}$ is an unknown $k$-vector of parameters; and $\sigma^2$ is an unknown positive parameter.   That is, $\bm{Y}$ follows a multivariate normal distribution with mean vector $\E(\bm{Y}) =\bm{\mu}=X\bm{\beta}$, for some $\bm{\beta}\in\Re^k$, and variance-covariance matrix $\var(\bm{Y})=\sigma^2\id$. The model (the set of possible vectors) for the mean vector is $\{\bm{\mu}=X\bm{\beta}: \bm{\beta}\in\Re^k\} = \sp(X)$. This is the {\itshape full model} in the discussion here.

The least-squares estimate of $\bm{\mu}=X\bm{\beta}$ in $\sp(X)$, which minimizes $(\bm{y}-X\bm{b})\pr(\bm{y}-X\bm{b})$, is $\hat{\bm{\mu}} = \P_X\bm{y}$.  The estimate $\hat{\bm{\mu}}$ of the mean vector is also called the vector of predicted values and denoted $\hat{\bm{y}}$.  A function $\hat{\bm{\beta}}$ of $\bm{y}$ is called a least-squares solution if and only if $X\hat{\bm{\beta}} = \P_X\bm{y}$ for all $n$-vectors $\bm{y}$. Residual, or error, sum of squares is $SSE = (\bm{y}-X\hat{\bm{\beta}})\pr(\bm{y}-X\hat{\bm{\beta}}) = \bm{y}\pr(\id-\P_X)\bm{y}$.  If $\bm{\mu}\in\sp(X)$, then $\hat{\sigma}^2 = MSE = SSE/\nu_E$, with $\nu_E = \tr(\id-\P_X)$, is an unbiased estimator of the population variance $\sigma^2$.  $MSE$ is mean squared error.

For a $k\times c$ matrix $G$, the function $G\pr\bm{\beta}$ is said to be {\itshape estimable} if and only if $\sp(G)\subset\sp(X\pr)$.  See Seely (1977) for a careful treatment of estimability and its relation to testing linear hypotheses of the form H$_0: G\pr\bm{\beta} = \bm{0}$, which is the subject of this discussion.  (Non-zero right-hand sides entail no essential complications, but we shall restrict attention here to $\bm{0}$ for simplicity.)

The conventional test statistic for a linear hypothesis takes the form of an $F$-statistic.  The denominator mean square is $MSE$.  The numerator sum of squares is a quadratic form $\bm{y}\pr P \bm{y}$, where $P$ is a symmetric, idempotent matrix such that $\sp(P)\subset\sp(X)$.  It follows that it is distributed as $\sigma^2$ times a chi-squared random variable with $\nu=\tr(P)$ degrees of freedom.  Its noncentrality parameter is $\delta^2_P = \bm{\beta}\pr X\pr PX\bm{\beta}/\sigma^2$.  

The fact that $\sp(P)\subset\sp(X)$ implies that $P\bm{y} = P\hat{\bm{y}}$, and so the numerator sum of squares is a function only of the estimated mean vector.  As a consequence, the numerator and denominator sums of squares are independent.

Let $N$ denote a matrix such that $\sp(N)=\{\bm{\beta}\in\Re^k: G\pr\bm{\beta}=\bm{0}\}$.  That is, $\sp(N)=\sp(G)^\perp$.  Under the condition that $G\pr\bm{\beta}=\bm{0}$, the restricted model is $\{X\bm{\beta}: \bm{\beta}\in\Re^k \text{ and } G\pr\bm{\beta}=\bm{0}\} = \sp(XN)$.  

Let $H$ denote a matrix with columns in $\sp(X)$ such that $X\pr H=G$.  Then $\sp(H) = \sp(X)\cap\sp(XN)^\perp$ and $\P_H = \P_X-\P_{XN}$.  (That $N\pr G=N\pr X\pr H = 0$ implies that $\sp(H)\subset\sp(X)\cap\sp(XN)^\perp$.  If $X\bm{b}\in\sp(X)\cap\sp(XN)^\perp$, then $N\pr X\pr X\bm{b}=\bm{0}$, which implies that $X\pr X\bm{b} = G\bm{c} = X\pr H\bm{c}$ for some $\bm{c}$, and hence $X\bm{b}-H\bm{c}$ is in $\sp(X)\cap\sp(X)^\perp = \{\bm{0}\}$, which implies that $X\bm{b}=H\bm{c}\in\sp(H)$.)

In order that the distribution of the $F$-statistic be central under the null hypothesis, $\delta_P^2 = 0$ if $G\pr\bm{\beta}=\bm{0}$. Because $P$ is nonnegative definite, that $\delta_P^2=0$ is equivalent to $\bm{\beta}\in\sp(X\pr P)^\perp$.  Thus a minimal requirement of $P$ is that $\sp(G)^\perp \subset\sp(X\pr P)^\perp$.  On the other hand, it is desirable that $\delta_P^2 > 0$ if $G\pr\bm{\beta}\neq \bm{0}$.  In order to satisfy both conditions, it must be true that $\sp(X\pr P)^\perp = \sp(G)^\perp$, which is equivalent to $\sp(X\pr P) = \sp(G)$. Proposition \ref{ns cond} in the appendix establishes that this is true if and only if $P=\P_H = \P_X - \P_{XN}$.  Subject to the condition that $\sp(P)\subset\sp(X)$, the only numerator sum of squares that yields an $F$-statistic that is central if and only if H$_0$ is true is $\bm{y}\pr \P_H\bm{y}$. In light of this, it is not surprising that the four methods described next all lead to the same numerator sum of squares.

The geometric heuristic addresses this question: given a vector $\bm{\mu}$ in $\sp(X)$, what criterion can be used to determine whether $\bm{\mu}$ is in the restricted model, that is, whether $\bm{\mu}\in\sp(XN)$? That is equivalent to whether $\bm{\mu}\pr (\P_X-\P_{XN})\bm{\mu}=0$.  At the same time, $\bm{\mu}\pr (\P_X-\P_{XN})\bm{\mu}$ is a squared distance function, and its magnitude gauges how far $\bm{\mu}$ is from the restricted model.  Substituting an estimate of $\bm{\mu}$, $\hat{\bm{\mu}} = X\hat{\bm{\beta}} = \P_X\bm{y}$, this results in the sum of squares $\bm{y}\pr(\P_X-\P_{XN})\bm{y}$.

The RMFM sum of squares is the difference in residual sum of squares for the restricted model and the full model.  It is
\[ \bm{y}\pr(\id-\P_{XN})\bm{y} - \bm{y}\pr(\id-\P_X)\bm{y} = \bm{y}\pr(\P_X - \P_{XN})\bm{y},
\]
the same as the geometric-method sum of squares.  The rationale behind it is that it measures how much restricting the model increases the lack of fit of the full model. 
It is customarily noted that this  sum of squares is one-to-one with the likelihood-ratio statistic.

Pearson (1900) illustrated chi-squared statistics in the form  \[\bm{u}\pr[\var(\bm{U})]^{-1}\bm{u}\], where $\bm{u}$ is the realization of the vector-valued random variable $\bm{U}$, designed to have expected value $\bm{0}$ under the hypothesis in question.  To test a hypothesis like H$_0: G\pr\bm{\beta}=\bm{0}$, such a statistic corresponds to a sum of squares like \[(G\pr\hat{\bm{\beta}})\pr[\var(G\pr\hat{\bm{\beta}})/\sigma^2]^-(G\pr\hat{\bm{\beta}}).\] 

Let $\hat{\bm{\beta}}$ be a least-squares solution.  With $H$ such that $\sp(H)\subset\sp(X)$ and $X\pr H = G$, $\P_X-\P_{XN} = \P_H$, as noted above, and hence $\bm{y}\pr (\P_X-\P_{XN})\bm{y} = \bm{y}\pr \P_H\bm{y}$.  Then
\begin{eqnarray*}
\bm{y}\pr \P_H\bm{y} &=& (H\pr\bm{y})\pr (H\pr H)^- (H\pr\bm{y})\\
&=& (G\pr\hat{\bm{\beta}})\pr [\var(G\pr\hat{\bm{\beta}})/\sigma^2]^-(G\pr\hat{\bm{\beta}})
\end{eqnarray*}
because, with $\P_XH=H$, $H\pr\bm{y} = H\pr X\hat{\bm{\beta}} = G\pr  \hat{\bm{\beta}}$ and $\var(H\pr\bm{Y}) = \sigma^2 H\pr H$.

At this point we have seen that the geometric sum of squares, the RMFM sum of squares, and the Pearson chi-squared sum of squares are identical.

The variance estimate approach is motivated by the balanced, single-classification ANOVA setting. With $n$ observations from each of $a$ populations (corresponding to ``treatments,'' say) all with population variance $\sigma^2$, the pooled within-sample mean square estimates $\sigma^2$ independent of any assumed relation among the population means.  The $a$ sample means all have variance $\sigma^2/n$.  Thus, if all the population means are the same, then $n$ times their sample variance also estimates $\sigma^2$. Their sample variance is Treatment Mean Square. 
This becomes the numerator mean square for the test statistic, with mean square within samples for the denominator.  Under the hypothesis of equal means, both mean squares estimate $\sigma^2$, and the rationale is that their ratio should be reasonably close to 1.  In fact, the ratio follows a central $F$ distribution if the $a$ population means are equal.

With unequal sample sizes, Treatment Sum of Squares becomes equivalent to Yates's (1934) $Q$, with $w_i=n_i$.  For the two-factor setting, Yates recognized that the same formulation could be applied to the ``marginal means of the subclass means'' because they are independent with variances $\sigma^2/w_i$ proportional to $\sigma^2$. Thus an estimator of $\sigma^2$ can be based on the marginal means and used as the numerator mean square.

This heuristic can be extended to the general setting as follows.  The objective is to devise a quadratic form in the estimated mean vector $\hat{\bm{y}}$ that is an unbiased estimator of $\sigma^2$ when H$_0$ is true.  That can be done directly as follows. Let columns of $V$ be an orthonormal basis for $\sp(X)$, so that $\P_X=VV\pr$, and let $\bm{Z}=V\pr\hat{\bm{Y}} =V\pr\bm{Y}$, because $\P_XV=V$.  Then $\bm{Z} \sim \N(V\pr X\bm{\beta}, \sigma^2\id)$.  Under H$_0: G\pr\bm{\beta}=\bm{0}$, $X\bm{\beta} \in\sp(XN)$, and the restricted-model estimate of $\sigma^2$ is residual mean square in this model, $\bm{z}\pr(\id-\P_{V\pr XN})\bm{z}$ divided by its degrees of freedom.  And 
\begin{eqnarray*} \bm{z}\pr(\id-\P_{V\pr XN})\bm{z}&=& \bm{y}\pr V(\id-\P_{V\pr XN})V\pr \bm{y}\\
&=& \bm{y}\pr (\P_X - \P_{XN})\bm{y}
\end{eqnarray*}
because $VV\pr=P_X$ and $V\P_{V\pr XN}V\pr = \P_{XN}$. 

While this development makes it clear that an estimator of $\sigma^2$ can be found in the restricted model for $\hat{\bm{Y}}$ when H$_0$ is true, and that the corresponding sum of squares is the same as the RMFM sum of squares, it does not parallel Yates's development for testing main effects in the two-factor analysis of variance model.  

To mimic Yates's construction, 
let $A$ and $C$ denote matrices such that $A$ has linearly independent columns in $\sp(X)$ and $X\pr AC = G$.  This guarantees that $D=A\pr A$ is positive definite (pd) and that $A\pr\bm{y} = (\P_XA)\pr\bm{y} = A\pr \P_X\bm{y} = A\pr\hat{\bm{y}}$ is a function of the estimated mean vector.  Entries of $\bm{U} = A\pr\bm{Y}$ correspond to Yates's marginal means of the subclass means.  
Matrices $A$ and $C$ satisfying these conditions exist in any case.  

With $\var(A\pr\bm{Y}) = D\sigma^2$, let $\bm{Z}=D^{-1/2}\bm{U} = D^{-1/2}A\pr\bm{Y}$, so that $\bm{Z}\sim \N(D^{-1/2}A\pr X\bm{\beta}, \sigma^2\id)$.  Let $M$ be a matrix such that $\sp(M)=\sp(C)^\perp$.  With the $c$ columns of $A$ linearly independent and in $\sp(X)$, it follows that 
\[\sp(A\pr X) = \sp(A\pr) = \Re^c.\]
 Then 
\begin{eqnarray*}\{A\pr X\bm{\beta}: \bm{\beta}\in\Re^k \text{ and } G\pr\bm{\beta}=\bm{0}\} &=& \{A\pr X\bm{\beta}: \bm{\beta}\in\Re^k \text{ and } C\pr A\pr X\bm{\beta}=\bm{0}\}\\
& =& \{\bm{\theta} \in\Re^c: C\pr\bm{\theta}=\bm{0}\}\\
& =& \sp(C)^\perp = \sp(M).
\end{eqnarray*}
If $\bm{\beta}$ is such that $G\pr\bm{\beta}=\bm{0}$ then $\bm{Z} \sim \N(D^{-1/2} M\bm{\gamma}, \sigma^2\id)$ for some $\bm{\gamma}$: this is the restricted model for $\bm{Z}$ under H$_0$.   Thus $MSE$ in this null model for $\bm{Z}$ is an unbiased estimator of $\sigma^2$.  Residual sum of squares in this model is
\begin{eqnarray} SSE_{\bm{z}} &=& \bm{z}\pr (\id-\P_{D^{-1/2}M})\bm{z}\nonumber\\
&=& \bm{u}\pr (D^{-1} - D^{-1}M(M\pr D^{-1}M)^-M\pr D^{-1})\bm{u}.\label{gen Yates}
\end{eqnarray}
This corresponds to (\ref{Q form}) and is equivalent to $Q$ in the setting that Yates (1934) considered, as shown in the next section.

Note further that
\begin{eqnarray*}
\bm{z}\pr(\id-\P_{D^{-1/2}M})\bm{z} &=& \bm{z}\pr \P_{D^{1/2}C}\bm{z}, \text{ by Prop. \ref{prop1}},\\
&=& \bm{y}\pr AC(C\pr D C)^-C\pr A\pr\bm{y} \\
&=& \bm{y}\pr \P_{AC}\bm{y}.
\end{eqnarray*}
With $X\pr AC = X\pr H = G$ and columns of both $AC$ and $H$ in $\sp(X)$, it follows that $AC=H$. Therefore $SSE_{\bm{z}} = \bm{y}\pr \P_H\bm{y} = \bm{y}\pr(\P_X-\P_{XN})\bm{y}$.

\section{Sum of Squares for A Main Effects in the Two-Factor ANOVA Model }
In the two-factor ANOVA model, denote levels of factors A and B by $i$ and $j$, respectively, $i=1,\ldots, a$, $j=1,\ldots, b$; denote the number of observations on the response under each factor-level combination (also called a {\itshape cell}) $i,j$ by $n_{ij}$, and assume that all $n_{ij}>0$ (there are no empty cells). Denote the population cell means of the response by $\eta_{ij}$ and the  $ab$-vector of cell means by $\bm{\eta}$. Let $n_{\cdot\cdot} = \sum_{ij}n_{ij}$.  For each observation $s=1,\ldots, n_{\cdot\cdot}$, define  the $s$-th row of the $n_{\cdot\cdot}\times ab$  matrix $\mbk$ to have  1 in the column corresponding to the factor-level combination $i,j$ under which the $s$-th subject was observed, and all other entries 0.  Then there is exactly one 1 in each row, and, in the $i,j$-th column, there are $n_{ij}$ 1s, $i=1,\ldots, a$, $j=1,\ldots, b$.  

Denote the $n_{\cdot\cdot}$-vector of the responses by $\bm{Y}$ and its realized value by $\bm{y}$.
The model for the mean vector $\bm{\mu}=\E(\bm{Y})$ of the response is $\mbk\bm{\eta}$, corresponding to $X\bm{\beta}$ in the general formulation above.  The columns of $\mbk$ are linearly independent.  

A consensus definition of A main effects (and how Yates (1934) defined them) is that they are differences among the A {\itshape population marginal means} $\bar{\eta}_{i\cdot} = (1/b)\sum_j\eta_{ij}$, $i=1,\ldots, a$. The $a$-vector of A marginal means can be expressed as $\bm{\theta}=(1/b)(\id_a\otimes \bm{1}_b)\pr\bm{\eta}$. 
The hypothesis of equal A marginal means is H$_0: S_a\bm{\theta}=\bm{0}$, or, in terms of $\bm{\eta}$, H$_0: [(1/b)(\id_a\otimes \bm{1}_b) S_a]\pr\bm{\eta}=\bm{0}$. This takes the form H$_0: G\pr\bm{\beta}=\bm{0}$ with $\bm{\beta}=\bm{\eta}$ and $G= (1/b)(\id_a\otimes \bm{1}_b) S_a = (1/b)(S_a\otimes \bm{1}_b)$.

Let $D_{ab}=(\mbk\pr \mbk)^{-1} = \Diag(1/n_{ij})$. 
To express the numerator sum of squares in the form (\ref{gen Yates}),  $X=\mbk$ and $G=X\pr AC$ with 
\[A=(1/b)\mbk D_{ab}(\id_a\otimes \bm{1}_b)\] 
and $C=S_a$. Then $M=\bm{1}_a$ so that $\sp(M)=\sp(C)^\perp$.
Then  $\bm{\theta}=A\pr X\bm{\beta} = A\pr \mbk\bm{\eta}=(1/b)(\id_a\otimes\bm{1}_b)\pr\bm{\eta}$ is the $a$-vector of A population marginal means $\bar{\eta}_{i\cdot}$; and 
$\hat{\bm{\theta}}=A\pr\bm{y}=(\bar{\bar{y}}_{i\cdot}=(1/b)\sum_j\bar{y}_{ij})$ is the $a$-vector of averages,  over levels of B, of the sample cell means $\bar{y}_{ij}=\sum_{\ell=1}^{n_{ij}}y_{ij\ell}/n_{ij}$ (which are the $ab$ entries in $\hat{\bm{\eta}}= D_{ab}\mbk\pr\bm{y}$).
Let 
\[ D_a = A\pr A = (1/b^2)(\id_a\otimes\bm{1}_b\pr)D_{ab}(\id_a\otimes\bm{1}_b) = (1/b^2)\Diag\left(\sum_j(1/n_{ij})\right).
\]
Diagonal entries of $D_a$ are $1/w_i$ in (\ref{Q form}).  With these specifications, 
(\ref{gen Yates}) 
is identical to (\ref{Q form}), the MWSM numerator sum of squares for A main effects. By the results in the last section, this is in turn equal to the other forms of the numerator sum of squares.  

Defining matrices $A$ and $C$ in this way corresponds to Yates's (1934) formulation. This has the consequence that $M=\bm{1}_a$  is a column vector, which avoids matrix operations in (\ref{gen Yates}). Another possible choice is $A=(1/b)\mbk D_{ab}$, so that $\bm{\theta}=\bm{\eta}$ and $C=S_a\otimes \bm{1}_b$.  That would result in $M$ having at least $ab-(a-1)$ columns.

The other forms are straightforward to re-express for this particular setting.  For $\bm{y}\pr\P_H\bm{y}$, for example, $H=\mbk D_{ab}(1/b)(S_a\otimes\bm{1}_b)$.     The full model is $\sp(\mbk)$.  A matrix $N$ such that $\sp(N)=\sp(G)^\perp = \sp(S_a\otimes \bm{1}_b)^\perp = \sp(\id_{ab}-S_a\otimes U_b)$ can be computed readily, so that the restricted model is $\sp(\mbk N)$.   Computation of any of these forms is quite straightforward.  Indeed, of all of them, Yates's expression appears to be the most complicated.

\begin{appendix}
\section{Notation, Definitions, and Facts}\label{defns}

In the notation shown next, assume for each that the matrix dimensions are such that the operations are defined.  Matrix notation is standard for addition, product, and inverse.  Generalized inverse and transpose of a matrix $A$ are denoted $A^-$ and $A\pr$, and $\tr(A)$ denotes the trace of $A$ if $A$ is square. Concatenation of columns of matrices $A$ and $B$ having the same number of rows is denoted $(A,B)$.

Vectors here are column vectors; they will be denoted in boldface, e.g., $\bm{z}$.  For an $n\times c$ matrix $M$, $\sp(M)$ denotes the linear subspace of real $n$-dimensional Euclidean space $\Re^n$ spanned by the columns of $M$: that is, $\sp(M) = \{M\bm{x}: \bm{x}\in\Re^c\}$. Orthogonality of vectors $\bm{u}$ and $\bm{v}$ in $\Re^n$ is defined by $\bm{u}\pr\bm{v}=0$.  
 The orthogonal complement of $\sp(M)$, denoted $\sp(M)^\perp$, is the set of all $n$-vectors that are orthogonal to all the vectors in $\sp(M)$.  
$\P_M$ denotes the orthogonal projection matrix onto $\sp(M)$: for any $n$-vector $\bm{z}$, $\P_M\bm{z}\in\sp(M)$ and $\bm{z}-\P_M\bm{z} \in \sp(M)^\perp$.  For any generalized inverse $(M\pr M)^-$ of $M\pr M$, $M(M\pr M)^-M\pr = \P_M$.
The relation between linear subspaces and their orthogonal projection matrices is one-to-one: $\sp(M_1)=\sp(M_2)$ if and only if $\P_{M_1} = \P_{M_2}$. Orthogonal projection matrices are symmetric and idempotent.

The orthogonal projection matrix $\P_M$ for a matrix $M$ can be computed as $BB\pr$, where columns of $B$ comprise an orthonormal basis for $\sp(M)$, which can be had by applying the Gram-Schmidt algorithm to $M$.  The expressions of the form $M(M\pr M)^- M\pr$ are used here to show relations among the several different forms of numerator sums of squares, not to suggest that computation of $\P_M$ requires a generalized inverse of $M\pr M$.  LaMotte (2014) shows, conversely, that a generalized inverse of $M\pr M$ can be had as a by-product of Gram-Schmidt on $M$. 

The Kronecker product of $A$ and $B$, denoted $A\otimes B$, is the matrix formed by replacing each entry $a_{ij}$ of $A$ by $a_{ij}B$.  It can be shown that $(A\otimes B)(C\otimes D) = (AC)\otimes(BD)$, $A\otimes (B+C) = A\otimes B + A\otimes C$, $(A\otimes B)\pr = A\pr \otimes B\pr$,  $\P_{A\otimes B} = \P_A\otimes \P_B$, and, if $A$ and $B$ are square, $\tr(A\otimes B) = \tr(A)\tr(B)$. If $A$ and $B$ have the same row dimension, then $(A,B)\otimes C = (A\otimes C, B\otimes C)$. Although $C\otimes (A,B)$ is not generally equal to $(C\otimes A, C\otimes B)$, the sets of columns are the same (in different order), and so $\sp[C\otimes(A,B)] = \sp(C\otimes A, C\otimes B)$.

  For a positive integer $m$, let $\bm{1}_m$ denote an $m$-vector of ones, $U_m = (1/m)\bm{1}_m\bm{1}_m\pr$, and $S_m = \id_m - U_m$.  For an $m$-vector $\bm{z}$, $U_m\bm{z}$ replaces each entry  in $\bm{z}$ by $\bar{z} = (1/m)\sum_i z_i$, and $S_m\bm{z}$ replaces each entry  $z_i$ by $z_i-\bar{z}$.  $S_m$ and $U_m$ are symmetric and idempotent, and $S_mU_m=0$.

  If $P$ is $n\times n$, symmetric, and idempotent, and if the $n$-variate random variable $\bm{Y}$ follows a multivariate normal distribution with mean vector $
E(\bm{Y})=\bm{\mu}$ and variance-covariance matrix $\sigma^2\id$ (signified as $\bm{Y} \sim \N_n(\bm{\mu}, \sigma^2\id_n))$, then $\bm{Y}\pr P\bm{Y}/\sigma^2 \sim \chi_\nu^2(\delta^2)$, with  $\nu=\tr(P)$ degrees of freedom and noncentrality parameter $\delta^2 = \bm{\mu}\pr P\bm{\mu}/\sigma^2$.

Proof of the following proposition is left to the reader.

\begin{propn}
Let $R$ be an $r\times c$ matrix, $M$ a matrix such that $\sp(M)=\sp(R)^\perp$, $D$ an $r\times r$ symmetric positive-definite (pd) matrix, $D^{1/2}$ a symmetric pd matrix such that $D^{1/2}D^{1/2}=D$, and $D^{-1/2}=(D^{1/2})^{-1}$. Then
\[  \P_{D^{1/2}R} = \id - \P_{D^{-1/2}M}.
\]\label{prop1}
\end{propn}

\begin{propn}
Let $X$ be an $n\times k$ matrix.  Let $G$ be a matrix such that $\sp(G)\subset\sp(X\pr)$, and let $N$ be a matrix such that $\sp(N)=\sp(G)^\perp$. If $P$ is a symmetric idempotent matrix such that $\sp(P)\subset\sp(X)$ , then $\sp(X\pr P)^\perp = \sp(G)^\perp$ iff $P=\P_X - \P_{XN}$.\label{ns cond}
\end{propn}

\vspace{.3cm}{\bfseries Proof.}  \fbox{$\imp$:} That $\sp(X\pr P)=\sp(N)^\perp$ $\imp$ $(XN)\pr P = 0$ $\imp$ $\P_{XN}P = 0$ $\imp$ $(\P_X-\P_{XN})P=\P_XP$, and, since $\sp(P)\subset\sp(X)$, $\P_XP=P$.  Therefore $\sp(P)\subset\sp(\P_X-\P_{XN})$.

That $\bm{z}\in\sp(\P_X-\P_{XN})$ $\imp$ $N\pr X\pr \bm{z} = \bm{0}$ $\imp$ $X\pr\bm{z}\in\sp(N)^\perp = \sp(X\pr P)$ $\imp$ $\exists$ $\bm{u}$ such that $X\pr\bm{z} = X\pr P\bm{u}$. With both $\bm{z}$ and $P\bm{u}$ in $\sp(X)$, this implies that $\bm{z}=P\bm{u} \in\sp(P)$. Therefore $\sp(\P_X-\P_{XN}) \subset\sp(P)$. Therefore $\sp(P) = \sp(\P_X-\P_{XN})$, and, because both $P$ and $\P_X-\P_{XN}$ are orthogonal projection matrices onto the same linear subspace, it follows that $P= \P_X -\P_{XN}$. 

\fbox{$\pmi$:} Suppose  $P=\P_X-\P_{XN}$.  
If $\bm{\beta}\in\sp(G)^\perp$ then $\exists$ $\bm{\gamma}$ such that $\bm{\beta}=N\bm{\gamma}$.  Then $PX\bm{\beta} = (\P_X-\P_{XN})XN\bm{\gamma} = \bm{0}$, which implies that $\sp(G)^\perp \subset\sp(X\pr P)^\perp$.  

If $\bm{\beta}\in\sp(X\pr P)^\perp$, then  $PX\bm{\beta} = \bm{0}$ $\imp$ $(\P_X-\P_{XN})X\bm{\beta} = \bm{0}$ $\imp$ $X\bm{\beta}=\P_{XN}X\bm{\beta} = XN\bm{\gamma}$ for some $\bm{\gamma}$. Because $\sp(G)\subset\sp(X\pr)$, $\exists$ $H$ such that $G=X\pr H$.  Then $G\pr\bm{\beta} = H\pr X\bm{\beta} = H\pr X N\bm{\gamma} = G\pr N\bm{\gamma} = \bm{0}$ because $\sp(N)=\sp(G)^\perp$.  Therefore $\sp(X\pr P)^\perp \subset\sp(G)^\perp$.  Therefore $\sp(G)^\perp = \sp(X\pr P)^\perp$.\hfill$\square$ 

\end{appendix}

\section{Bibliography}
\begin{itemize}\setlength{\topsep}{0cm}\setlength{\labelsep}{0cm}
\setlength{\itemsep}{0cm}\setlength{\parsep}{0cm}
\setlength{\parskip}{0cm}\setlength{\itemindent}{-1cm}
\item[] Anderson, R. L., and Bancroft, T. A. (1952).  Statistical Theory in Research.  McGraw-Hill Book Company, New York.
\item[] Hector, A., von Felten, S., and Schmid, B. (2010).  Analysis of variance with unbalanced data: an update for ecology \& evolution.  Journal of Animal Ecology 79: 308-316.
\item[] Herr, D. G. (1986).  On the history of ANOVA in unbalanced, factorial designs: the first 30 years.  The American Statistician 40: 265-270.
\item[] Hocking, R. R. (2013). Methods and Applications of Linear Models, Third Edition.  John Wiley \& Sons, Inc., Hoboken, New Jersey.
\item[] Khuri, A. I. (2010). Linear Model Methodology.  Chapman \& Hall/CRC, Boca Raton, FL.
\item[] LaMotte, L. R. (2014). The Gram-Schmidt construction as a basis for linear models.  The American Statistician 68: 52-55.
\item[] Langsrud, \O. (2003).  ANOVA for unbalanced data: Use Type II instead of Type III sums of squares.  Statistics and Computing 13:163-167.
\item[] Macnaughton, D. B. (1998).  Which sums of squares are best in unbalanced analysis of variance? MatStat Research Consulting Inc.
\item[] Pearson, K. (1900). On the criterion that a given set of deviations from the probable in the case of a correlated system of variables is such that it can be reasonably supposed to have arisen from random sampling. Philosophical Magazine, Series 5, 50: 157-172.
\item[] SAS Institute Inc. (1978).  SAS Technical Report R-101, Tests of hypotheses in fixed-effects linear models.  Cary, NC.
\item[] Searle, S. R. (1971).  Linear Models.  John Wiley \& Sons, Inc., New York.
\item[] Searle, S. R. (1987).  Linear Models for Unbalanced Data.  John Wiley \& Sons, Inc., New York.
\item[] Searle, S. R. (1994).  Analysis of variance computing package output for unbalanced data from fised-effects models with nested factors.  The American Statistician 48: 148-153.
\item[] Searle, S. R., Speed, F. M., and Henderson, H. V. (1981). Some computational and model equivalences in analyses of variance of unequal-subclass-numbers data.  The American Statistician 35: 16-33.
\item[] Seely, J. (1977). Estimability and linear hypotheses. The American Statistician 31: 121-123.
\item[] Smith, C. E., and Cribbie, R.  (2014).  Factorial ANOVA with unbalanced data: A fresh look at the types of sums of squares.  Journal of Data Science 12: 385-404.
\item[] Snedecor, G. W., and Cox, G. M. (1935). Disproportionate subclass numbers in tables of multiple classification.  Research Bulletin No. 180, Agricultural Experiment Station, Iowa State College of Agriculture and Mechanic Arts, Ames, Iowa.
\item[] Yates, F. (1933). The principles of orthogonality and confounding in replicated experiments.  The Journal of Agricultural Science 23:108-145.
\item[] Yates, F. (1934).  The analysis of multiple classificatioins with unequal numbers in the different classes.  Journal of the American Statistical Association, 29(185): 51-66.
\end{itemize}

\end{document}